\documentclass[aps,showpacs,floatfix,a4paper,pra,preprint,groupedaddress]
{revtex4}

\usepackage[T1]{fontenc}
\usepackage[english]{babel}
\usepackage{amsmath}
\usepackage{graphicx}
\usepackage{calc}
\graphicspath{{./}{Figs/}}
\usepackage{bm,bbm}

\newcommand{\bess}[2]{\mbox{$J_{#1}(#2)$}}
\newcommand{\mexp}[1]{\mbox{$\mathrm{e}^{#1}$}}

\newcommand{\mRe}{\mathrm{Re}}
\newcommand{\mIm}{\mathrm{Im}}

\begin{document}
\title{ Larmor frequency dressing by a non-harmonic transverse magnetic field}
\author{G. Bevilacqua}
\author{V. Biancalana}

\author{Y. Dancheva}
\author{L. Moi}
\affiliation{ CNISM, CSC and Dipartimento di Fisica, Universit\`a di Siena,
  via Roma 56, 53100 Siena, Italy}

\begin{abstract}
We present a theoretical and experimental study of spin precession in
the presence of both a static and an orthogonal oscillating magnetic
field, which is non-resonant, not harmonically related to the Larmor
precession and of arbitrary strength. Due to the intrinsic
non-linearity of the system, previous models that account only for 
the simple sinusoidal 
case cannot be applied. We suggest an alternative approach and develop
a  model that  closely agrees  with experimental  data produced  by an
optical-pumping   atomic   magnetometer.   We  demonstrate   that   an
appropriately designed non-harmonic field makes it possible to extract
a linear  response to a weak  dc transverse field,  despite the scalar
nature  of the  magnetometer,  which normally  causes  a much  weaker,
second-order response.
\end{abstract}

\date{\today}


\maketitle

\section{Introduction}
The precession of spins in a  static homogeneous field is a well known
problem  of classic  as  well  as quantum  physics.   A second  widely
studied system  is derived from  the previous one when  an oscillating
field  perpendicular  to  the  static   one  is  added.  This  is  the
configuration used in typical magnetic resonance (MR) experiments, and
for  this reason,  it has  been  extensively studied.   In MR  setups,
however, the transverse field  oscillates at a frequency near-resonant
to the  Larmor frequency set by  the static field, and  has usually an
amplitude much  smaller than the  static one.  The  resonant condition
greatly facilitates  the task of modeling the  system: the oscillating
field is seen as a  superposition of two counter-rotating (and usually
weak) fields, one of which appears to be (quasi-) static in a reference
frame which rotates  at the Larmor frequency around  the static field,
while  the  other  has  negligible  effects.  This  is  well-known  in
textbooks as the rotating wave approximation (RWA).

In this  work we study the  system described above,  but in conditions
which make the RWA unsuitable. In particular, we consider the cases in
which  the transverse  field has  a general  periodic  time dependence
containing  Fourier components  at  frequencies much  larger than  the
Larmor frequency.  In fact,  we re-examine, extend, and
generalize  the   results  reported  decades  ago   by  S.Haroche  and
co-workers  \cite{CCT_1970}, where similarly  an analytical  model has
been developed and applied to interpret experimental data from optical
detection of atomic precession.

The results  presented here are  modeled with the Larmor  equation and
discussed using a perturbative  approach based on the Magnus expansion
of the  time-development operator. Compared to previous  works on this
subject, our approach makes it easy  to model the effects of a transverse
field oscillating with an arbitrary waveform.

It seems  worth stressing  that the subject  has implications  for the
research    area   of   dynamic    localization   \cite{Kenkre_art186,
  Holtaus_1996}, which concerns the effect of a time-periodic field on
quantum    systems   such   as    charged   particles    in   crystals
\cite{Kenkre_art81}      and       Bose      Einstein      condensates
\cite{PhysRevA.79.013611}.

The  experimental implementation  is  feasible, provided  that a  slow
precession  occurs  around a  weak  static  field,  for instance  with
ultra-low-field NMR setups and  in atomic spin precession experiments.
This  latter is the  nature of  the setup  used in  our case,  and the
reported experimental  results concern measurements  performed with an
optical-pumping atomic magnetometer operated  at a $\mu$T static field
range.  The corresponding Larmor frequency is in the kHz range, making
it easy  to apply a  homogeneous transverse field oscillating  at much
higher frequencies, even with a much larger amplitude.

Our  apparatus  detects the  time-dependent  Faraday  rotation of  the
polarization  plane of  a weak  probe  laser beam,  which is  directly
related to a component of the macroscopic magnetization of atoms.  The
analyzed magnetization component is  the one parallel to the direction
of  the  oscillating  field,  which  coincides  with  the  probe  beam
propagation axis.  In the  presence of a strong transverse oscillating
field the three components of  the magnetization evolve in time with a
complex   behaviour,  which   is   far  from   a  simple   precession.
Nevertheless the  measured component  maintains its similarity  to the
simple precession  behaviour.  In other  terms, the motion  around the
direction ($z$)  of the  static field is  deeply affected by  a strong
non-resonantly oscillating field along  $x$, which mainly modifies the
dynamics of  the magnetization components $M_y$ and  $M_z$, but leaves
$M_x$  evolving  (approximately) with  a  simple  harmonic law.   This
quasi-harmonic  evolution  of $M_x$  makes  it  possible  to define  a
modified (dressed) Larmor frequency even when the oscillating field is
so  strong that  no precession  motion could  be recognized  by simply
observing the time evolution of the whole $\mathbf{M}$ vector.

The       paper      is       organized      as       follows:      in
Section~\ref{sec:experimental-setup}  we   describe  the  experimental
setup  used,  in Section~\ref{sec:model}  we  discuss the  theoretical
model,  and  Section~\ref{sec:disc} contains  the  discussion and  the
conclusions.

\section{Experimental setup \label{sec:experimental-setup}}

\begin{figure}[htbp]
   \centering
  \includegraphics{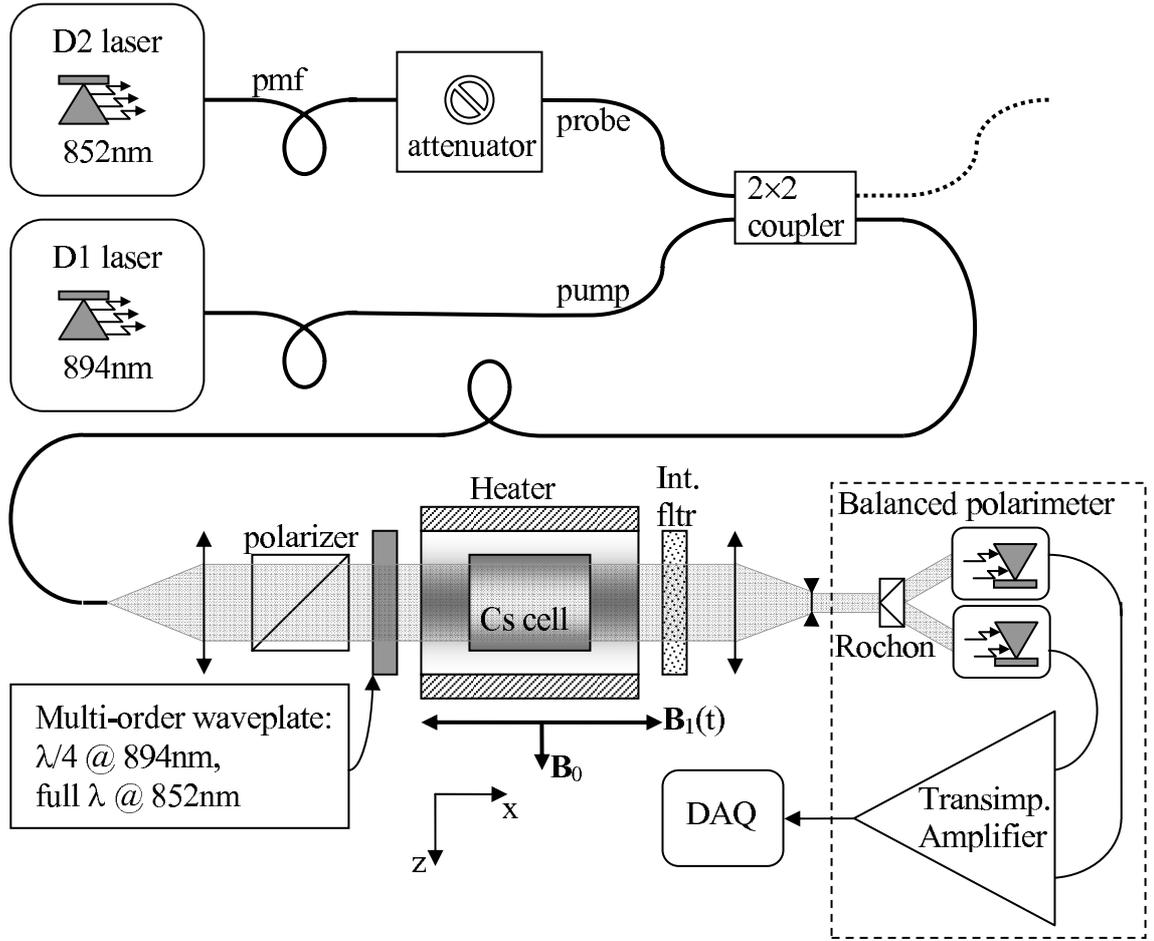}
  \caption{Schematics of  the experimental  setup. The probe  laser is
    tuned to the D2-Cs line  @852nm and unmodulated; the pump laser is
    tuned to the D1-Cs line @894nm and is modulated synchronously with
    the atomic  precession. Both are  mixed and coupled to  the sensor
    though   polarization  maintaining  fibres   (pmf).  The   probe  is
    attenuated down  to $\mu$W level before  the Cs cell,  the pump is
    stopped by an  interferential filter (Int fltr) after  the cell. A
    multi-order waveplate renders the pump polarization circular, while
    leaving  the probe polarization  linear. The  Cs cell  contains 23
    Torr of N$_2$ as a buffer gas, which quenches the fluorescence and
    slows down  the Cs atom motion,  making it diffusive.  The cell is
    slightly  heated to  increase Cs  vapour density.  The transmitted
    probe  laser polarization  is analyzed  by a  balanced polarimeter
    made of a Rochon polarizer  oriented at $45^o$ with respect to the
    probe  polarization  axis  and  two  photodiodes.  A  differential
    transimpedance  amplifier converts the  photocurrent to  a voltage
    signal, which  is then digitally acquired.  That signal reproduces
    the Faraday rotation, associated to the time-dependent orientation
    of the precessing atomic spins.}
  \label{fig:appsp}
\end{figure}

The basic  structure of the experimental  setup is that  of an optical
atomic  magnetometer (see  Fig.  \ref{fig:appsp}).   Using  a resonant
circularly  polarized  laser  light,  alkali  atoms are  pumped  in  a
specific Zeeman sublevel  and thus the medium is  magnetized along the
wave-vector  ($x$)  direction.   The  atomic  medium is  placed  in  a
homogeneous  magnetic field  ($z$  direction) transverse  to the  pump
light propagation.   Consequently the Zeeman sublevel  where atoms are
pumped evolves in time, giving rise to the magnetization precession in
the $xy$ plane, around the static magnetic field. 

This precession  is detected  by analyzing the  Faraday rotation  of a
weak,  linearly polarized,  probe beam  which propagates  in  the $xy$
plane (in our case it is collinear to the pump beam). The linear probe
beam polarization  can be decomposed in two  $\sigma^+$ and $\sigma^-$
counter-rotating  circularly  polarized  components, which  experience
variation of the respective refractive  indexes due to the atomic spin
precession.   These   two  circular   components   recombine  with   a
time-dependent relative dephasing in  a linear polarization rotated by
an angle  oscillating at the  Larmor frequency.  Beside  the different
dispersions,  the  $\sigma^+$ and  $\sigma^-$  components suffer  also
different absorptions, but this aspect  is made of little relevance by
tuning the probe laser away from the resonance center.

To contrast the relaxation processes that cause a progressive decay of
the induced magnetization, and hence  of the detected signal, the pump
light is  applied periodically,  synchronously with the  precession of
the atomic spins, i.e. resonantly with the Larmor frequency set by the
static magnetic field.

In  our setup,  the  pump  radiation is  generated  by a  single-mode,
distributed  feedback,  pigtailed   DFB  diode  laser,  whose  optical
frequency  is   controlled  through   the  junction  current   and  is
periodically made resonant  (or non resonant) to the  D1 transition of
Cs, at  894~nm. It  illuminates the atomic  sample at an  intensity of
about  0.5~mW/cm$^2$.    The  probe   radiation  is  generated   by  a
single-mode, Fabry-Perot, pig-tailed diode  laser resonant with the D2
line  of Cs,  at  852~nm. It  illuminates  the sample  at much  weaker
intensity, being attenuated at a level of about half $\mu$W/cm$^2$, in
order to negligibly perturb the sample.

Polarization maintaining  (PM) fibres bring  the radiation of  the two
lasers to  a coupler (via an  on-fibre attenuator in the  case of the
probe) which  mixes them in the  PM fibre illuminating  the sample. In
proximity of the  atomic sample, the two radiations  emerging from the
fibre  end are  collimated by  a  lens, their  linear polarization  is
reinforced by a polarizing cube.Then, a phase-plate acts differently on the
polarization  of both  the  radiations, rendering  the pump  radiation
circular, while leaving  the probe one linear.  This  double effect is
achieved thanks to a multi-order wave plate made of a 480$\mu$m quartz
layer,  which  introduces  \cite{1999_ghosh}  for the  crossed  linear
polarization components of the two wavelengths a (5+1/4)$\lambda_{D1}$
and a  5$\lambda_{D2}$ relative  delays, respectively. Such  method of
tailored multi-order waveplate method is quite similar to that applied
in  the  magnetometric  setup   (based  on  Rb  vapors)  described  in
\cite{johnson2010}.

After  interacting  with the  atomic  sample,  the  pump radiation  is
blocked by an interferential filter and the probe beam polarization is
analyzed by  a balanced polarimeter  made of a 45$^o$  oriented Rochon
and  a  couple  of  photodetectors whose  photocurrent  difference  is
amplified by a low-noise transimpedance amplifier.

The  polarimetric signal  is  either  analyzed as  a  function of  the
laser-modulation  frequency   to  characterize  the   atomic  magnetic
resonance, or used to generate  a signal for driving the laser optical
frequency,   thus  closing   a   loop  which   makes   the  system   a
self-oscillator.   In  the first  case,  the  precession frequency  is
estimated with  a best fit procedure,  while in the second  case it is
evaluated as  the self-oscillation  frequency, provided that  the loop
delay  is   properly  compensated.   In   practice,  the  closed-loop,
self-oscillating operation  helps in shortening  the measurement time,
but may introduce systematic errors due to inappropriate choice of the
loop  delay.  On the  contrary,  operating  in  the open  loop/scanned
frequency/best fit  procedure requires a  longer time but  improves the
accuracy  in the  Larmor  frequency determination.  All the  plots
reported  in  Sec.~\ref{sec:disc}   were  obtained  in  this  more
accurate  open-loop  operation  mode.   With  the  used  static  field
intensities, the  precession frequency ranges  from hundreds of  Hz to
several kHz.

The  atomic  sample  is  made  of a  high  quality  gas-buffered  cell
containing a  droplet of Cs and 23  Torr of N$_2$.  The  buffer gas is
suitable for  slowing down the thermal  motion of Cs  atoms, making it
diffusive and  also quenches  any stray fluorescence  light. Following
several optimization  campaigns devoted to improve  the sensitivity of
the  setup  in magnetometric  applications,  the  pump  laser is  made
resonant  (at the  right  time) with  the  $|^1S_{1/2}, F_g=4  \rangle
\rightarrow |^1P_{1/2}  \rangle$ hyperfine components of  the D1 line,
while the  analyzed ground state is  the $F_g=4$ and  the detection is
made tuning the probe laser about 2\  GHz on the blue wing of the line
made   of   the  triplet   $|^1S_{1/2},   F_g=4  \rangle   \rightarrow
|^2P_{3/2}\rangle$.  The  pump radiation  is broadly modulated  with a
slightly   asymmetric  signal,  so   that,  besides   the  synchronous
excitation of  the $|^1S_{1/2}, F_g=4  \rangle$ level, it  also spends
some   time  in   resonance  with   the  $|^1S_{1/2},   F_g=3  \rangle
\rightarrow| ^2P_{3/2} \rangle $ triplet.

The Cs cell is inserted in an anti-inductive heating coil by which the
temperature is  increased up  to about 50$^{o}$C  in order to  reach a
good compromise for  having a large signal (an  optimal optical depth)
without  introducing an excessive  broadening due  to the  increase of
spin-exchange collision  rate. A second coaxial,  inductive coil makes
it  possible to apply  a homogeneous  time dependent  transverse field
which  is thus  parallel to  the optical  axis.  This  second  coil is
supplied by  a digital waveform generator which  provides an arbitrary
periodic voltage  signal. This signal  is designed to  produce current
(and  hence  $x$-oriented  magnetic  field)  Fourier  components  with
appropriate  phases  and amplitudes,  in  accordance  to the  measured
complex impedance of the coil.

The  whole experiment  is conducted  in a  magnetically  clean volume,
where  a set  of three  mutually perpendicular,  large  size Helmholtz
coils and a  set of quadrupoles make it possible  to control the three
components of the magnetic  field and reduce their inhomogeneities. In
particular, the  horizontal components ($x$,  $y$) of the  Earth field
are  fully  compensated  while   the  vertical  one  ($z$),  partially
compensated down to a few $\mu$T, determines the static field.

The  system  is  here  used  for an  absolute  frequency  measurement.
Nevertheless,  the  apparatus  is  designed  to  perform  differential
measurements as well. In fact, the fibre coupler is a $2\times 2$ 50\%
device, meaning that the two  radiations are mixed in equal ratios and
made  available in  two output  PM  fibres.  Thus,  the apparatus  can
easily be adapted to work with a dual sensor, suitable for gradiometry
and  for  differential  magnetic  measurements. The  accuracy  in  the
absolute frequency  measurement (order of  1 Hz) is largely  in excess
compared  to  the   needs  of  this  work,  and   is  limited  by  the
environmental magnetic noise.

\section{\label{sec:model} Model}
The aim of this section is to describe and characterize the precession
of  spins  in  the  presence   of  a  time  dependent  magnetic  field
orthogonally oriented with respect to  the static one, using a special
form of the perturbation theory for the time evolution operator. 

As stated above, taking the $z$ axis in the direction of the
static   field  and   the  $x$   axis  along   the   alternating  one,
i.e.  $\mathbf{B}  =  (B_x\,  \phi(\omega  t),  0,  B_z)$,  the  Larmor
equations for the magnetization are
\begin{equation}
  \label{eq:La:ini}
  \begin{cases}
    \dot{M}_x & = - \Omega_z M_y  \\
    \dot{M}_y & =  \Omega_z M_x -\Omega_x {\phi}(\omega t)\, M_z\\
    \dot{M}_z & =  \Omega_x {\phi}(\omega t)\, M_y
  \end{cases}
\end{equation}
Here $\Omega_{x,z}  = \gamma B_{x,z}$  ($\gamma $ is  the gyromagnetic
ratio) and  $\phi(\omega t)$  is a real  and periodic function  with a
well-behaved   Fourier   expansion.    Notice   that   hereafter   the
magnetization is  treated classically, but the same  results hold true
if the $M_i, \; i=x,y,z$ are treated as non-commuting quantum mechanical
operators.

Using  the dimensionless time  $\tau  = \omega  t$,  the equation  set
(\ref{eq:La:ini}) can be recast as
\begin{equation}
  \label{eq:La:adi}
  \frac{d}{d \tau} \mathbf{M} =
  \left[
  \frac{\Omega_z}{\omega}
  \begin{pmatrix}
    0 & -1 & 0 \\
    1 & 0 & 0\\
    0 & 0 & 0
  \end{pmatrix}  
  +
  \frac{\Omega_x}{\omega}
  \begin{pmatrix}
    0 & 0 & 0 \\
    0 & 0 & -1\\
    0 & 1 & 0
  \end{pmatrix}
  \phi(\tau)
\right] \mathbf{M} .
\end{equation}
We are interested in the case of $\zeta \equiv \Omega_z/\omega \ll 1$,
so it  is natural to  use perturbation theory.   To keep track  of the
perturbation  order, we  introduce  the parameter  $\epsilon$ and  let
$\zeta  \rightarrow  \epsilon  \zeta$  in the  intermediate  formulas,
we  set then $\epsilon=1$ at the  end of the  calculations.  We can
thus rewrite eq.~\eqref{eq:La:adi} as
\begin{equation}
  \label{eq:La:pert:eps}
    \frac{d}{d \tau} \mathbf{M} =
  \left[ \epsilon  \,\zeta\, A_z +
    \xi \, \phi(\tau) A_x \right] \mathbf{M},
\end{equation}
where   $\xi \equiv  \Omega_x/\omega$.   Notice   that  introducing   the
commutator  $A_y  \equiv  [A_z,  A_x]$,  the  three  matrices  $A_k,\; k=x,y,z$ constitute
essentially the  three-dimensional real representation  of the angular
momentum quantum operators.

The magnetization  $\mathbf{M}(\tau)$ can be obtained  acting with the
time evolution operator $U(\tau)$
   as  $   \mathbf{M}(\tau)   =
U(\tau)\mathbf{M}(0)$ on the initial value $\mathbf{M}(0)$.
In the spirit of perturbation theory, it is useful to factorize $U$ in
the ``interaction'' representation as
\begin{equation}
  \label{eq:U:fact}
  U(\tau) = \exp\left[ \xi \int_0^{\tau} \phi(\tau') d\tau'
    \, A_x\right]\; U_I(\tau) =
  \begin{pmatrix}
    1 & 0 & 0 \\
    0 & \cos \Phi & -\sin \Phi \\
    0 &\sin \Phi & \cos \Phi
  \end{pmatrix} \; U_I(\tau)
\end{equation}
which implicitly defines $U_I(\tau)$. Moreover we set 
\begin{equation}
  \label{eq:def:PHI}
\Phi(\tau) =  \xi \int_0^{\tau} \phi(\tau') d\tau'.
\end{equation}
Let's  comment on  the  structure of  eq.\eqref{eq:U:fact}: the  first
factor  represents a  rotation of  an  angle $\Phi$  around the  first
axis.  This rotation  relates  the laboratory  frame  to the  rotating
frame, thus $\Phi$  is the Larmor angle for  the precession around the
oscillating  field.  In the  rotating  frame,  the evolution  operator
satisfies
\begin{equation}
  \label{eq:UI}
  \frac{d U_I}{d \tau} = \epsilon  
  \zeta
  \mexp{-\Phi(\tau) A_x} A_z \mexp{\Phi(\tau) A_x} \; U_I
= \epsilon  
  \zeta \left[ \cos \Phi \, A_z + \sin \Phi \, A_y \right] \; U_I
= \epsilon \zeta A_z^{(I)}\; U_I
\end{equation}
whose      solution,      following      the     Magnus      expansion
device~\cite{2009_Blanes}, can be written as
\begin{equation}
   \label{eq:UI:magnus}
   U_I(\tau) = \mexp{W(\tau)}.
 \end{equation}
 The exponent satisfies the non-linear equation \cite{2009_Blanes}
 \begin{equation}
   \label{eq:exp:magnus}
   \frac{dW}{d\tau} = \epsilon \zeta\, A_z^{(I)}
   - \frac{1}{2} \left[W, \epsilon \zeta\, A_z^{(I)}\right]
   + \frac{1}{12} \left[W, \left[W, \epsilon  \zeta\,A_z^{(I)}\right] \right]
   + \ldots
 \end{equation}
 which can be solved perturbatively.
In  fact expanding $W$ as  
$ W = \sum_{n\geq  1}   \epsilon^n  \,  W_n$  one  gets   the  first  order
 contribution
\begin{equation}
  \label{eq:W1:contr}
  W_1(\tau) = \zeta \, \mRe(\Psi(\tau)) \, A_z + \zeta \, \mIm (\Psi(\tau)) \, A_y
\end{equation}
where we introduced the quantity
\begin{equation}
  \label{eq:int:expifi}
  |\Psi(\tau)| \mexp{i \beta(\tau)}=
  \int_0^{\tau} \mexp{i\,\Phi(\tau')}\, \mathrm{d} \tau'.
\end{equation}
After some algebra one finds
\begin{equation}
  \label{eq:form:UI}
  U_I = \begin{pmatrix}
    \cos (\zeta|\Psi|) &
    -\cos \beta\, \sin (\zeta|\Psi|) &  
    \sin \beta\, \sin (\zeta|\Psi|)\\
    \cos \beta\, \sin (\zeta|\Psi|) &
    \cos^2 \beta \cos (\zeta|\Psi|) +\sin^2\beta &
    \sin (2\beta) \sin^2(\zeta|\Psi|/2) \\
    -\sin \beta\, \sin (\zeta|\Psi|) &
    \sin (2\beta) \sin^2(\zeta|\Psi|/2)  &
    \sin^2 \beta \cos (\zeta|\Psi|) +\cos^2\beta
    \end{pmatrix}
\end{equation}
which represents a rotation of an angle $\zeta|\Psi|$ around the axis
\begin{equation}
  \label{eq:axis:UI}
  \mathbf{n} =
  \begin{pmatrix}
    0,&  \sin \beta,& \cos\beta
  \end{pmatrix}.
\end{equation}
Considering the initial magnetization prepared along the $x$ axis
$\mathbf{M}(0) = \left( M_x(0), 0, 0 \right)$
and using the eqs.~\eqref{eq:form:UI} and~\eqref{eq:U:fact} one finds
\begin{equation}
  \label{eq:Mx:final}
  M_x(\tau) = M_x(0) \, \cos( \zeta |\Psi(\tau)| ),
\end{equation}
clarifying  that   $\Psi$  is  the   Larmor  angle  around   the  axis
$\mathbf{n}$.   Deriving  this   quantity,  the   first-order  correct
frequency of the signal experimentally analyzed can be evaluated
\begin{equation}
  \label{eq:WL:1ord}
  \Omega^{(1)}_L  =  \omega  \frac{d}{d \tau}\,  \left(  \zeta|\Psi(\tau)|
  \right).
\end{equation}
Notice that, on general ground, if  one needs only the argument of the
cosine  in \eqref{eq:Mx:final},  it  can be  more  easily obtained  by
calculating the eigenvalues of $W_1$.

Another case of practical interest is the following
\begin{equation}
  \label{eq:La:pert:eps2}
    \frac{d}{d \tau} \mathbf{M} =
  \left[ \epsilon \,\zeta\, A_z + \epsilon^2 \, \xi_0\, A_x +
    \xi \, \phi(\tau) A_x \right] \mathbf{M},
\end{equation}
which models the case of a very small (residual) static magnetic field
superimposed  to the  alternating one.  Precisely we  assume  that the
dimensionless quantities $\xi$, $\zeta$ and $\xi_0$ satisfy
\begin{equation}
  \label{eq:assump:eps2}
  \xi \equiv \gamma B_x^{\mathrm{(ac)}}/\omega \approx O(1)
\qquad \zeta \equiv\epsilon \,\gamma B_z/\omega \ll 1 \qquad
\xi_0 \equiv \epsilon^2\,\gamma B_x^{\mathrm{(dc)}}/\omega \ll \zeta,
\end{equation}
showing the need of  second-order perturbation theory to  study the effect
of $\xi_0$. Repeating  the above derivation we get  the same $W_1$ while
the second-order contribution $W_2$ becomes
\begin{equation}
  \label{eq:W2:eps2}
  W_2(\tau) = \left[ \xi_0 \tau -
\frac{\zeta^2}{2}\mIm \int_0^\tau \Psi(\tau')\mexp{-i\Phi(\tau') \mathrm{d}
  \tau'}\right] A_x \equiv \chi(\tau)\, A_x.
\end{equation}
The  geometric  meaning  of  $\chi$   is  that  in  the  second  order
approximation the  rotation described in  eq. \eqref{eq:W1:contr} does
not occur  around $\mathbf{n}$, but  around a direction having  also a
small $x$  component.  Finding the  eigenvalues of $ \epsilon\,  W_1 +
\epsilon^2 W_2 $ makes it 
possible  to determine the  argument of the
cosine
\begin{equation}
  \label{eq:WL:2ord:angolo}
  \Theta_L(\tau) = \epsilon \sqrt{\zeta^2 |\Psi(\tau)|^2 + \epsilon^2 \, \chi(\tau)^2}
\end{equation}
and thus eq. \eqref{eq:WL:1ord} generalizes to
\begin{equation}
  \label{eq:WL:2ord}
  \Omega^{(2)}_L = \omega \frac{d \Theta_L(\tau)}{d \tau}.
\end{equation}

\subsection{\label{sec:determ-psi}Determination   of  \protect{$\Psi$}
  and \protect{$\chi$}}
We   are  interested  in    periodic  signals,   so  that
$\phi(\tau)$, $\Phi(\tau)$ and  $\exp ( i \Phi) $  have a well behaved
Fourier expansion. More precisely we consider signals $\phi(\tau)$ with a zero
mean so that
\begin{equation}
  \label{eq:exp:IF:expa}
  \mexp{i \, \Phi(\tau)} = \sum_{n=-\infty}^{\infty} G_n
  \mexp{i\,n\,\tau} =
  \sum_{n=-\infty}^{\infty} |G_n|
  \mexp{i( \,n\,\tau + \theta_n)},
\end{equation}
then it is easy to obtain
\begin{equation}
  \label{eq:exp:IF:int}
  \int_0^\tau \mexp{i \, \Phi(\tau')} \mathrm{d} \tau'
  = G_0 \tau + \sum_{n \neq 0} \frac{G_n}{i n} \left(
    \mexp{i\,n\,\tau} - 1 \right)
= G_0 \tau + g(\tau).
\end{equation}
where the  implicitly defined function $g(\tau)$  is a limited  and oscillating
function.
When  $\tau \gg 1 $ (or $t \gg  1/\omega$) we can neglect
$g(\tau)$ obtaining $  \Psi(\tau) = G_0 \tau$,
and thus we find a dressed Larmor frequency
\begin{equation}
  \label{eq:dress:larmor}
  \Omega^{(1)}_L = \Omega_z \left| G_0 \right|.
\end{equation}

To access the second-order corrections we need the integral
\begin{equation}
  \label{eq:int:chi}
  \int_0^{\tau} \Psi(\tau')  \mexp{-i\,\Phi(\tau')} \mathrm{d} \tau' =
   |G_0|^2 \tau^2/2 + \tau i ( a - b ) + f_0(\tau) + \tau f_1(\tau)
\end{equation}
where $f_0(\tau)$  and $f_1(\tau)$ are periodic  and limited functions
whose explicit form is not important here. The parameters $a$ and $b$ are
defined as
\begin{equation}
   \label{eq:alfa:beta}
  a \equiv \sum_{n \neq 0} \frac{G_0^*G_n}{n} \qquad
  b \equiv \sum_{n \neq 0} \frac{|G_n|^2}{n}.
\end{equation}
Neglecting the oscillating terms in eq. \eqref{eq:int:chi} as before and
taking the imaginary part we get
\begin{equation}
  \label{eq:chi:fin}
  \chi(\tau) = \tau \left[ \xi_0 +
    \zeta^2  \left(  b  -   \mRe(a) \right)/2 \right] \equiv
  \tau \left( \xi_0 + \zeta^2 c /2 \right) \equiv \tau \chi_0.
\end{equation}

The dressed Larmor frequency  eq.~\eqref{eq:WL:2ord} can thus be evaluated
approximating the square root in eq.~\eqref{eq:WL:2ord:angolo} as
\begin{equation}
  \label{eq:WL:1e2}
  \frac{   \Omega^{(2)}_L}{\Omega_z}   \approx   |G_0|  \left[   1   +
    \frac{\chi_0^2}{2 \zeta^2 |G_0|^2} \right]
  = |G_0| + \frac{1}{2 |G_0|}
  \left[ (\xi_0/\zeta)^2 + \xi_0 c + \zeta^2c^2/4 
  \right],
\end{equation}
where  now  we can  consider  the last  terms  as  corrections to  the
first-order   result    \eqref{eq:dress:larmor}.    Notice   how   the
coefficients $G_n,\;\, n\neq 0$ are encoded in the $c$ parameter.

\subsection{\label{sec:pure-sinusoidal}Pure sinusoidal signal}
This    case   has    already   been    studied   and    reported   in
literature.  S.~Haroche  et  al.,  in  a  pioneer  paper  had  already
demonstrated the possibility of solving  the simple case of a harmonic
transverse field  ~\cite{CCT_1970} with a different  approach based on
quantum  mechanics.   In  that   study,  the  authors   quantized  the
oscillating field  as a  bosonic mode (rf  photons) and  used standard
first-order perturbation theory to  evaluate the energy shifts.  Then,
in the (semi-classical)  limit of a large boson  number, they obtained
the  same result  reported  here  below. That  method  would run  into
difficulties for  generic time-periodic  signals.  On the  other hand,
let us examine how quickly the results are obtained using our
approach. Starting from $\phi(\tau) = \cos \tau$
we easily get $ \Phi(\tau) = \xi \sin \tau$
and the coefficients $G_n$ are the Bessel functions \cite{abramowitz:stegun}
\begin{equation}
  \label{eq:Gn:CCT}
  G_n = \left| J_n\left( \xi \right) \right|
  \mexp{i \theta_n}
\end{equation}
being the phases $\theta_n$ either $0$ or $\pi$ according to the sign
of $J_n$. We thus re-obtain the Haroche et al. result
\begin{equation}
  \label{eq:CCT:gen}
  \Omega_L = \Omega_z\,
\left|J_0\left( \xi   \right) \right|.
\end{equation}

Notice that in case of a general harmonic signal
$\phi(\tau)  = \cos(
\tau - \tau_0)$ one has
\begin{equation}
  \label{eq:harm:gen:int}
  \Phi(\tau) = \xi
  \left[ \sin(\tau -\tau_0 ) + \sin\tau_0    
  \right]
\end{equation}
and thus
\begin{equation}
  \label{eq:Gn:gen:harm}
  G_n = \mexp{i\,\xi \sin\tau_0}\,\left| J_n\left( \xi \right) \right|
  \mexp{i (\theta_n - n \tau_0)}
\end{equation}
giving for $|G_0|$ the same result as before.

\subsection{Non-harmonic signals \label{sec:anharmonic-signal}}

Having  found  the  general expression~\eqref{eq:dress:larmor}  for  the
dressed Larmor  frequency, we can study more  complicated cases. Let's
first introduce a higher  harmonic component in the signal, indicating
with $r$ the ratio between its amplitude and the fundamental one i.e.
\begin{equation}
  \label{eq:anha:gen}
  \phi(\tau) = \cos\tau + r \cos(k\tau + \varphi)
\end{equation}
which gives
\begin{equation}
  \label{eq:anha:int}
  \Phi(\tau) = \xi
  \left[ \sin \tau + \frac{r}{k}\left( \sin(k \tau + \varphi) -
    \sin\varphi\right)
    \right].
\end{equation}

Repeating the same steps as before, one gets
\begin{equation}
  \label{eq:Gn:anha:gen}
  G_n = \mexp{-i\,(\xi r/k) \sin \varphi} \,
  \sum_{p=-\infty}^{+\infty}
  J_{n-kp}\left( \xi \right)\,
  J_{p}\left( \xi \,r /k \right)\,
  \mexp{i\,p\,\varphi}
\end{equation}
which is  a generalized Bessel  function~\cite{Dattoli1996}.  Thus the
leading   contribution (eq.~\eqref{eq:dress:larmor} )  of   the   Larmor
frequency becomes
\begin{equation}
  \label{eq:larmor:ana:Omega}
  \frac{ \Omega_L}{ \Omega_z} = F(\xi,r,k,\varphi) \equiv \left|
    \sum_{p=-\infty}^{+\infty}
    J_{-kp}\left( \xi \right)\,
  J_{p}\left( \xi \,r/k \right)\,
  \mexp{i\,p\,\varphi}
\right|.
\end{equation}

Another highly non-harmonic signal which gives analytical results is the square
wave
\begin{equation}
  \label{eq:sq:wave:signal}
    \Phi(\tau) = \xi 
    \begin{cases}
      \tau & 0 \leq \tau \leq 2 \pi \alpha\\
      \alpha(2\pi - \tau)/(1-\alpha) & 2\pi\alpha < \tau < 2\pi
    \end{cases}.
\end{equation}
In     fact    in     this    case     the     integration    outlined
in eq.~\eqref{eq:exp:IF:expa} can be done trivially obtaining
\begin{equation}
\label{G0:sq_wave}
G_0 =
 \mexp{i\pi \xi \alpha}\, \frac{ \sin(\pi \xi \alpha) }{ \pi \xi \alpha},
\end{equation}
and thus the dressed Larmor frequency results
\begin{equation}
  \label{eq:WL:sw}
  \frac{\Omega_L }{\Omega_z} = \frac{|\sin \pi \xi \alpha |}{\pi \xi \alpha}.
\end{equation}

For  general periodic signals  the $G_n$  coefficients can  be obtained
analytically  as   multiple  sums  of  products   of  ordinary  Bessel
functions.   However,  this  approach  soon becomes  cumbersome  and  a
numerical evaluation of the $G_n$, using for instance the Fast Fourier
Transform, is a viable alternative.

\section{Discussion}
\label{sec:disc}

The main issue of the paper  is to study the effect of an alternating,
non-resonant magnetic field oriented perpendicularly to the static one
causing  a reduction of the precession rate of spins.  Such a
reduction  has  a non-monotone  dependence  on  the  amplitude of  the
alternating  field,  and, under  specific  conditions, the  precession
frequency can take a zero value.  This phenomenon has been
previously  observed and theoretically  interpreted in  the case  of a
harmonic signal \cite{CCT_1970}.

The model  introduced in this work  is suitable to  the description of
the simple case of a  harmonically oscillating field but also makes it
possible to easily extend the analysis to the non-harmonic case.

The comparison of theoretical  predictions and experimental results is
made  taking into  account  several operating  conditions and  varying
different parameters. We preliminarily test the consistency with known
results obtained in the simple  case of a transverse field oscillating
with a  harmonic law. Then we  test the strongly non-harmonic  case of a
square-wave field. A more detailed  analysis is presented for the case
of a  two-Fourier-component field,  where we analyse  the effect  as a
function  of the  amplitudes  and of  the  relative phase  of the  two
components. We also  devote attention to the cases  when the frequency
ratio  is even  ($k=2$) or  odd ($k=3$),  pointing out  the symmetries
discussed   in   the   appendix  \ref{sec:app:prop:F}.    Finally   we
demonstrate the  existence of  a linear response  to a weak  dc field,
which  has implications  in the  design of  atomic  magnetometers with
vectorial response.

\begin{figure}[htbp]
  \centering
  \includegraphics[width=8cm]{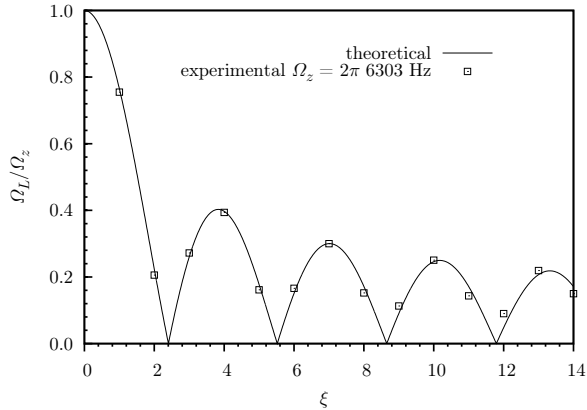}
  \caption{  Ratio  between  the  dressed and  unperturbed  precession
    frequency as a function of  the amplitude of the oscillating field
    (expressed in the dimensionless quantity $\xi=\gamma B_x /\omega$)
    in  the  case  of   a  harmonic  transverse  field,  according  to
    eq.~\eqref{eq:CCT:gen}  the  plot   reproduces  the  behaviour  of
    $\left|J_0\left( \xi \right) \right|$.  }
  \label{fig:CCTlike}
\end{figure}

We show in Fig.~\ref{fig:CCTlike}  the comparison of experimental data
obtained       with       the       apparatus       described       in
Sec.~\ref{sec:experimental-setup} and  the curve (namely  $\left | J_0
\right |$) given by the model
in the case of a harmonic  field.  In this figure (as in the following
ones) the amplitude of the  oscillating field is expressed in terms of
the   dimensionless   quantity    $\xi$   defined   (see   also   Sec.
\ref{sec:model} eq.~\eqref{eq:La:pert:eps}) by $\xi=\Omega_x/\omega=\gamma B_x /\omega$.  The
frequency  reduction effect  is  also given  in natural  dimensionless
units,  reporting the ratio  between the  dressed and  the unperturbed
precession frequencies.  It is worth  noticing that if the same curves
are plotted  in terms absolute  quantities (i.e. plotting  the dressed
precession frequency as a function  of $B_x$), passing from a spinning
species to  another having a larger gyromagnetic  factor, the vertical
axis scales  up while  the horizontal one  scales down. This  makes it
possible to  identify conditions  for matching the  dressed precession
frequencies  of different  species,  which is  of  great interest  for
possible  applications, see  for example  \cite{ito03}  and references
therein.

\begin{figure}[htbp]
  \centering
  \includegraphics[width=8cm]{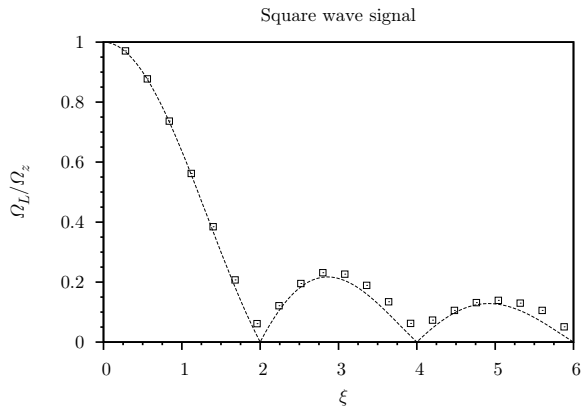} 
  \caption{   Comparison  of   theoretical   predictions  (line)   and
    experimental  points  for  the   ratio  between  the  dressed  and
    unperturbed precession frequency as a function of the amplitude of
    a    square-wave   like    oscillating    field.   According    to
    eq.~\eqref{eq:WL:sw} the behaviour is described by the modulus of a
    sinc function.}
  \label{fig:sw:comp}
\end{figure}

As a first  example of a non-harmonic modulation,  we consider the case
of a symmetric  (50\% duty cycle) square wave. In  this case the model
predicts     a     behaviour    given     by     a    sinc     modulus
(eq.~\eqref{eq:sq:wave:signal}).   The prediction  is in  good agreement
with   the   experimental   observation,   as   appears   evident   in
Fig.~\ref{fig:sw:comp}.

\begin{figure}[htbp]
  \centering
  \includegraphics[width=8cm]{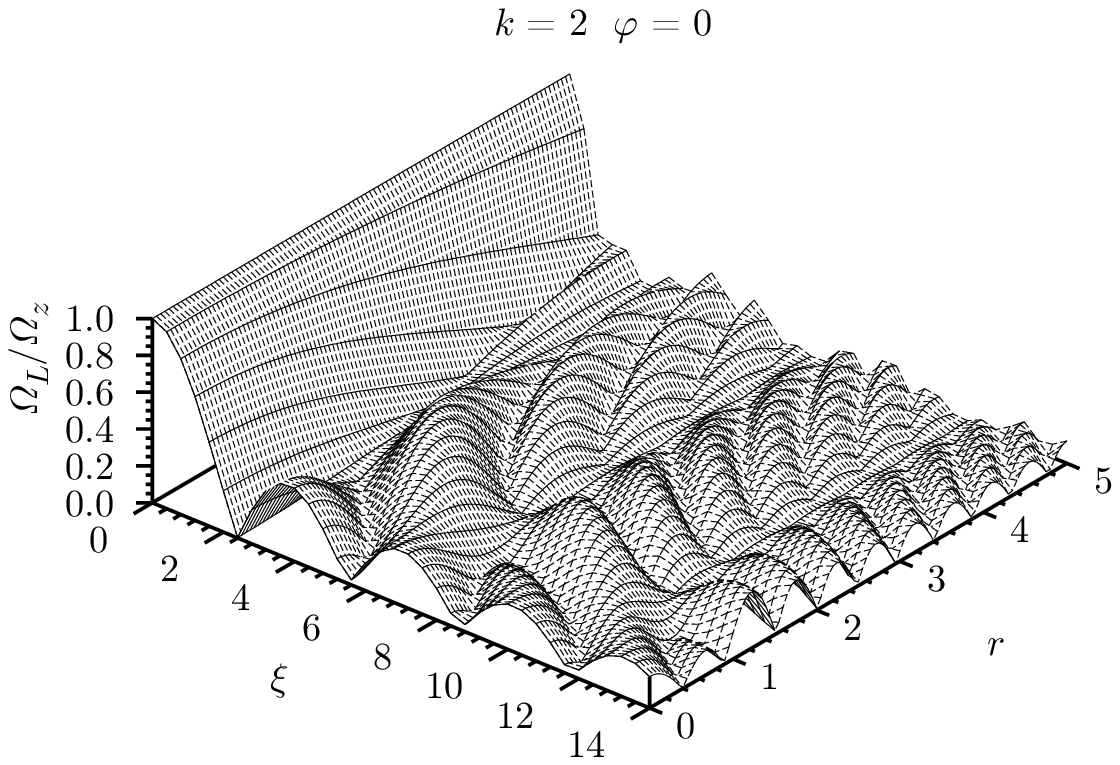}
  \includegraphics[width=8cm]{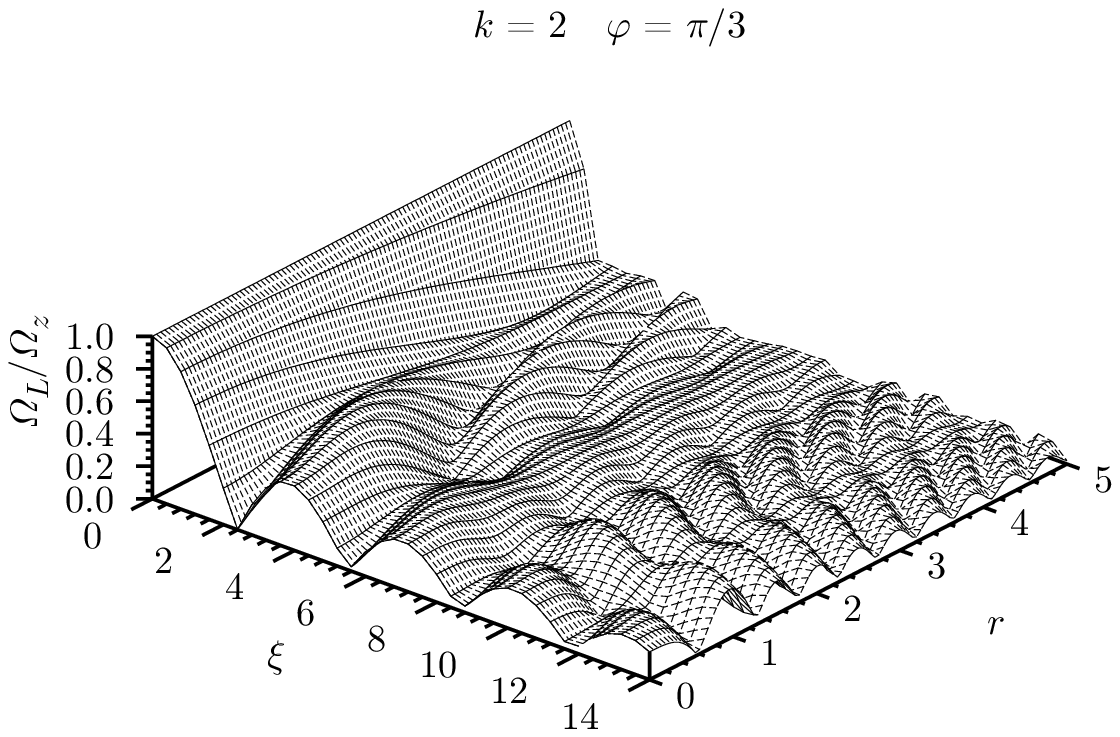}\\
  \includegraphics[width=8cm]{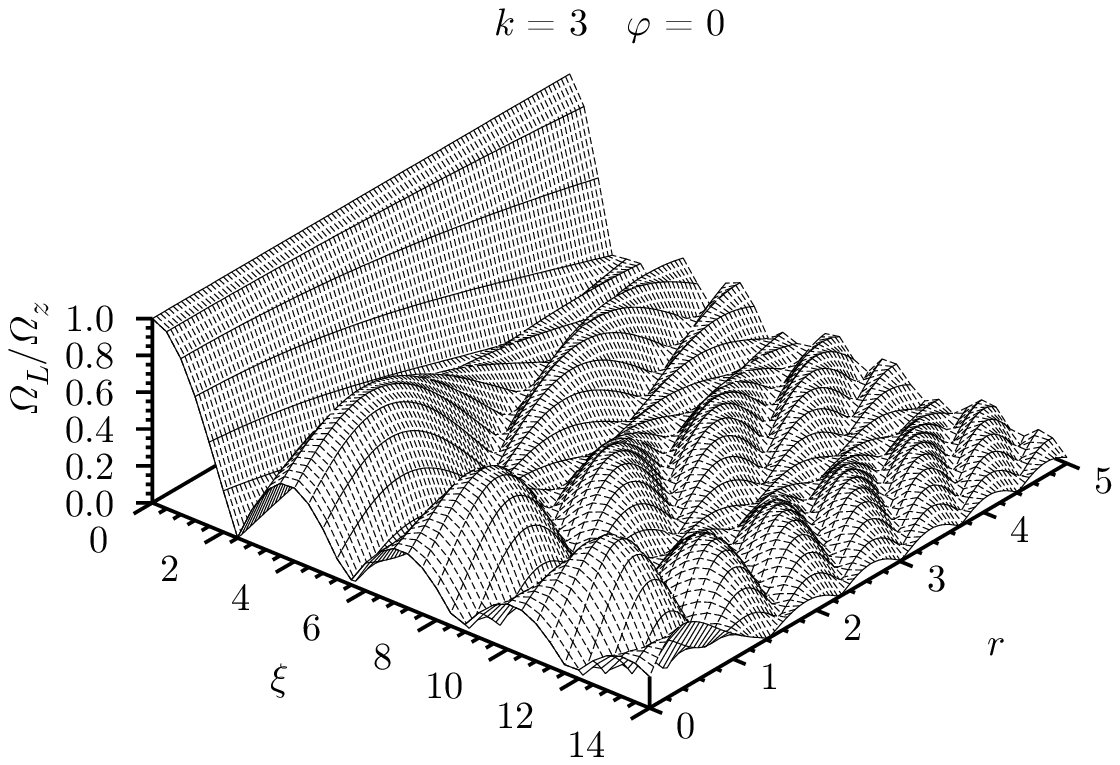}
  \includegraphics[width=8cm]{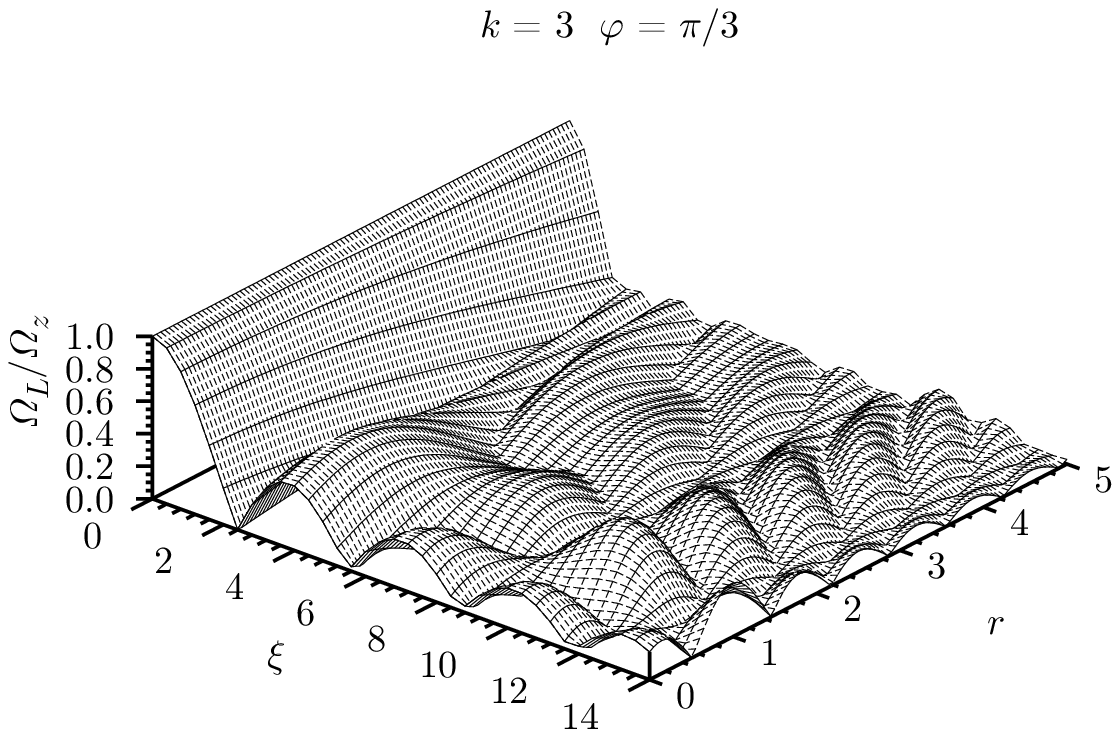}
  \caption{
    Graphs of the function $F(\xi,r,k,\varphi)$, 
    see eq.~\eqref{eq:larmor:ana:Omega}, for different values
    of $k$  and $\varphi$. 
  }
  \label{fig:F}
\end{figure}

A detailed analysis is performed in the case of a non-harmonic periodic
transverse field  made of only  two Fourier components  (a fundamental
one plus  its $k^{th}$  harmonics). In this  case the  prediction (see
eq.~\eqref{eq:larmor:ana:Omega})  can   be  easily  compared   to  the
experimental results when varying one of the independent variables, as
in  the  following.  In  Fig.~\ref{fig:F}  we  preliminarily show  the
behaviour  of  $F$  as  a  function  of the  amplitude  $\xi$  of  the
fundamental  Fourier component  and  the relative  amplitude $r$  (see
eq.~\eqref{eq:anha:gen}  for  the  precise  meaning of  $r$,  $k$  and
$\varphi$)  of  the $k^{th}$  harmonics,  for  fixed  values of  their
relative phase and of $k$.

\begin{figure}[htbp]
  \centering
  \includegraphics[width=8cm]{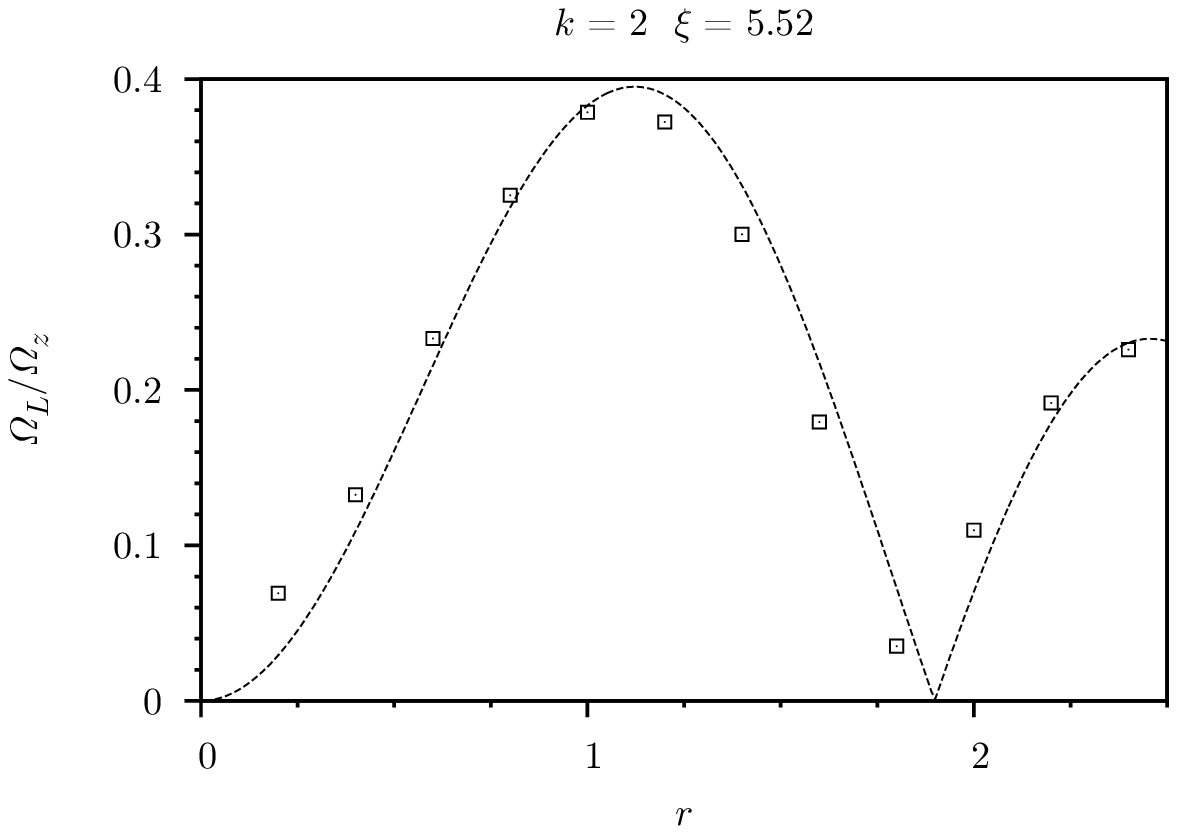}\\
  \includegraphics[width=8cm]{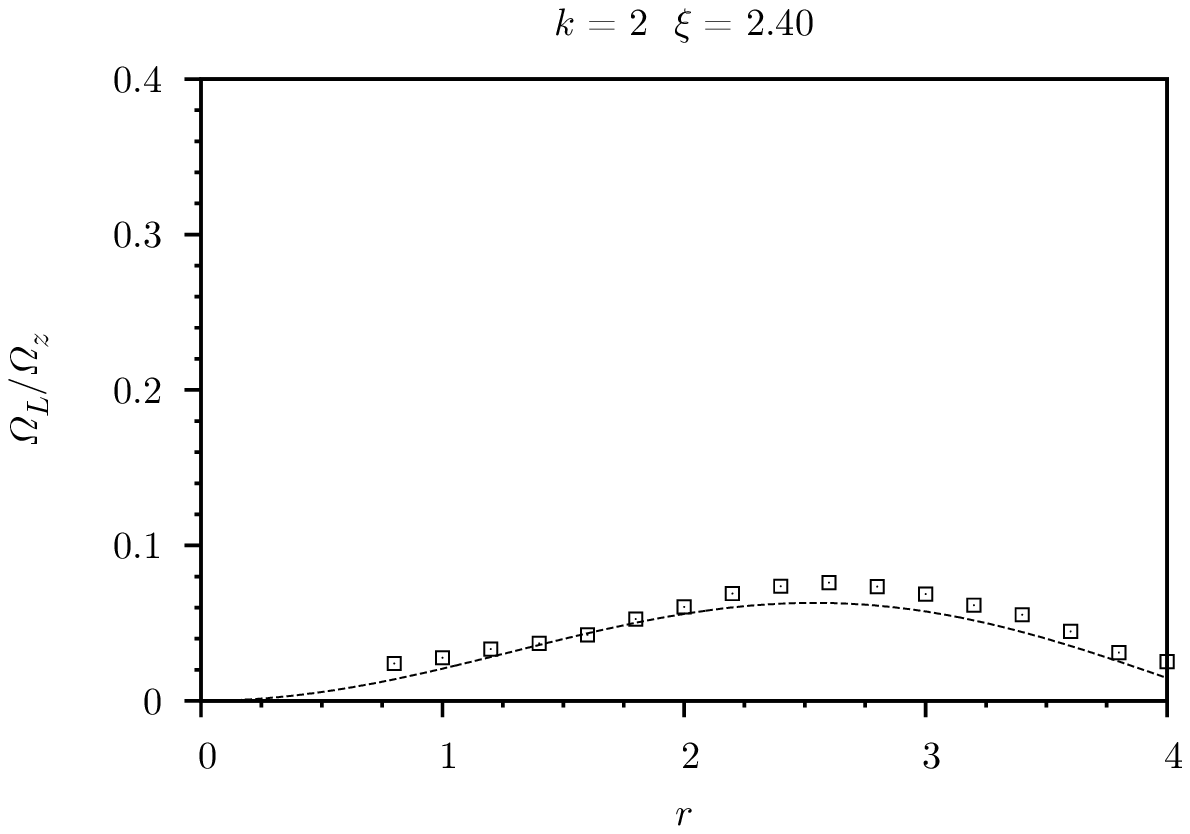}\\
  \includegraphics[width=8cm]{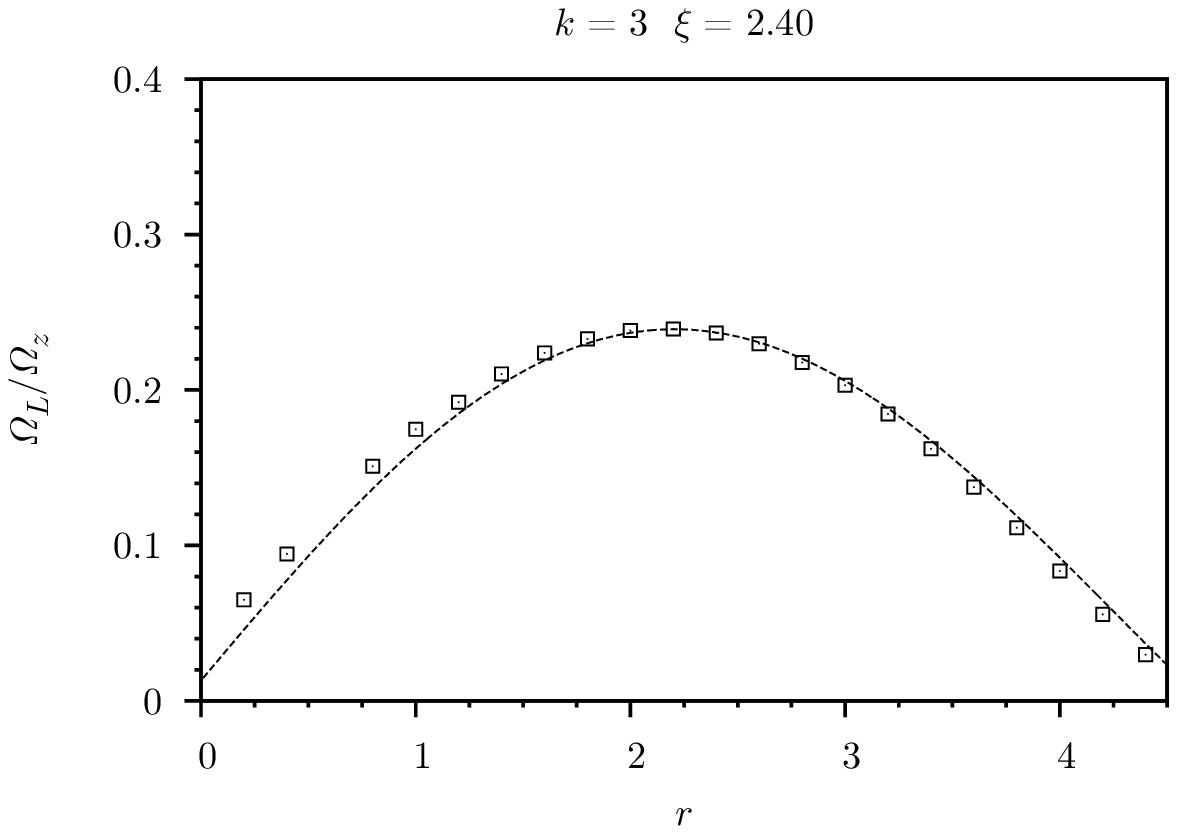}
  \caption{Comparison   of    theoretical   predictions   (line)   and
    experimental  points for  different values  of  $\xi$ and  $k$ as  a
    function of $r$ (see eq.~\eqref{eq:anha:gen}). }
  \label{fig:scan_r}
\end{figure}

Due to  technical limitations, only  a few lobes  of the ones  shown in
Fig.~\ref{fig:F} have been accessed to compare the model prediction
with the experimental results. As an example, in Fig.~\ref{fig:scan_r}
the  dependence of  $F$  on the  amplitude  ratio $r$  is presented  for
several couples $(k, \xi)$ of the frequency ratio and of the fundamental
amplitude.

\begin{figure}[htbp]
  \centering
  \includegraphics[width=8cm]{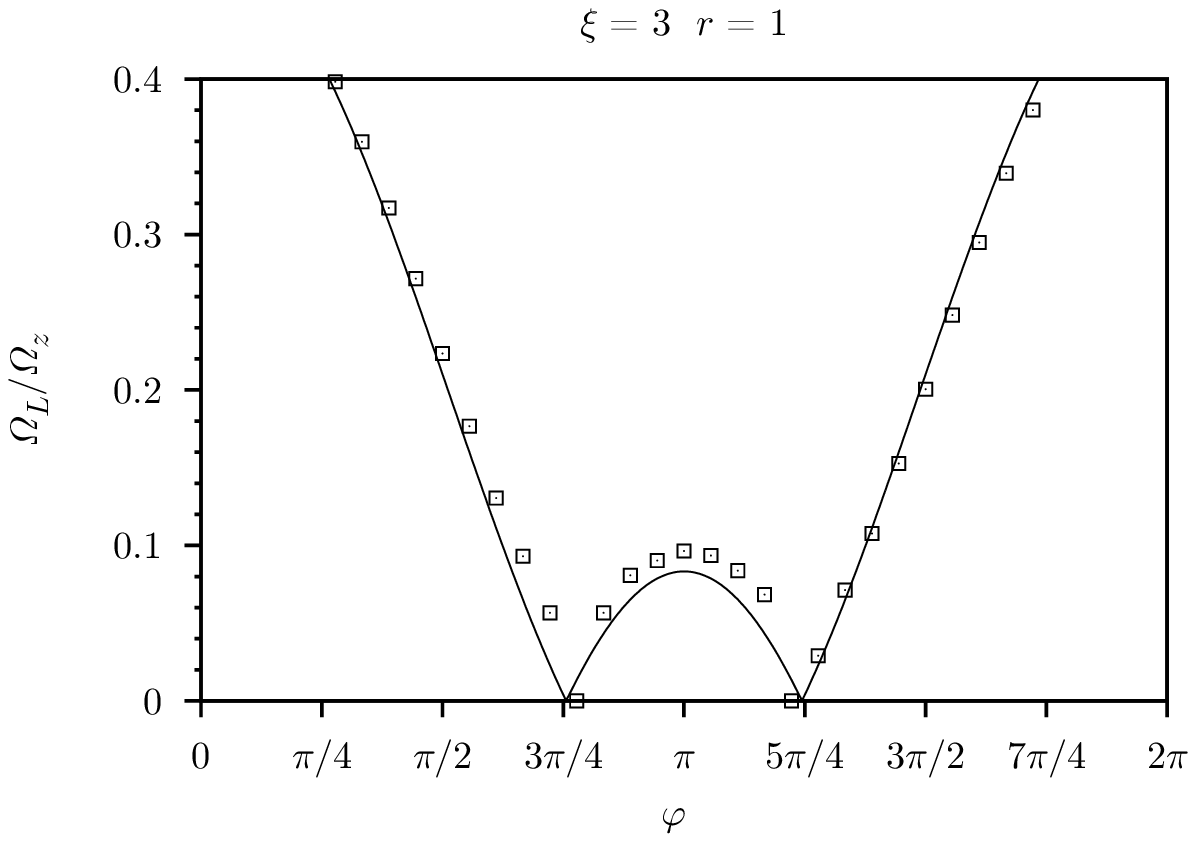}\\
  \includegraphics[width=8cm]{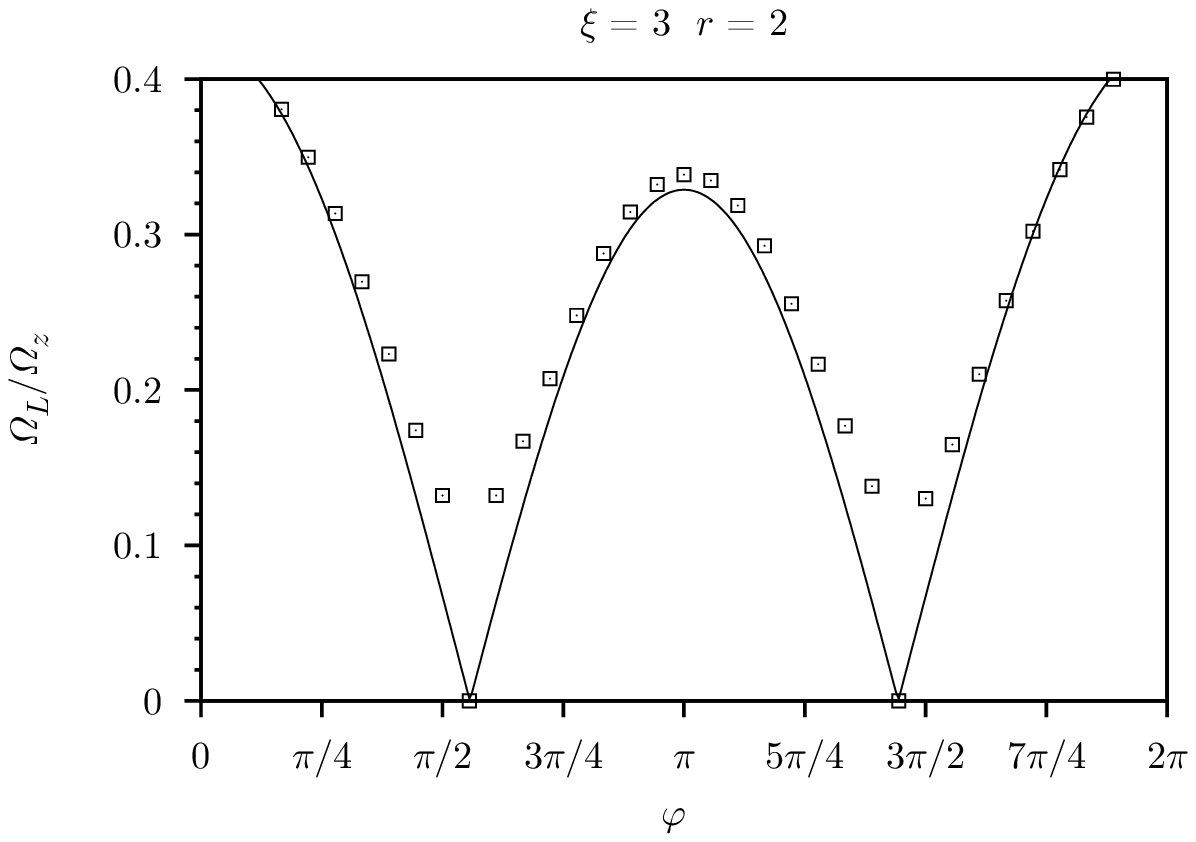}
  \caption{   Comparison  of   theoretical   predictions  (line)   and
    experimental points for $k=3$, $\xi=3$ and  two  values of  and $r$
    for  the  ratio between  the  dressed  and unperturbed  precession
    frequencies  as a  function of  the  relative phase  between the  two
    Fourier components of the oscillating field (see eq.~\eqref{eq:anha:gen}). }
  \label{fig:k3:x3x3_r1r2_comp}
\end{figure}

\begin{figure}[htbp]
  \centering
  \includegraphics[width=8cm]{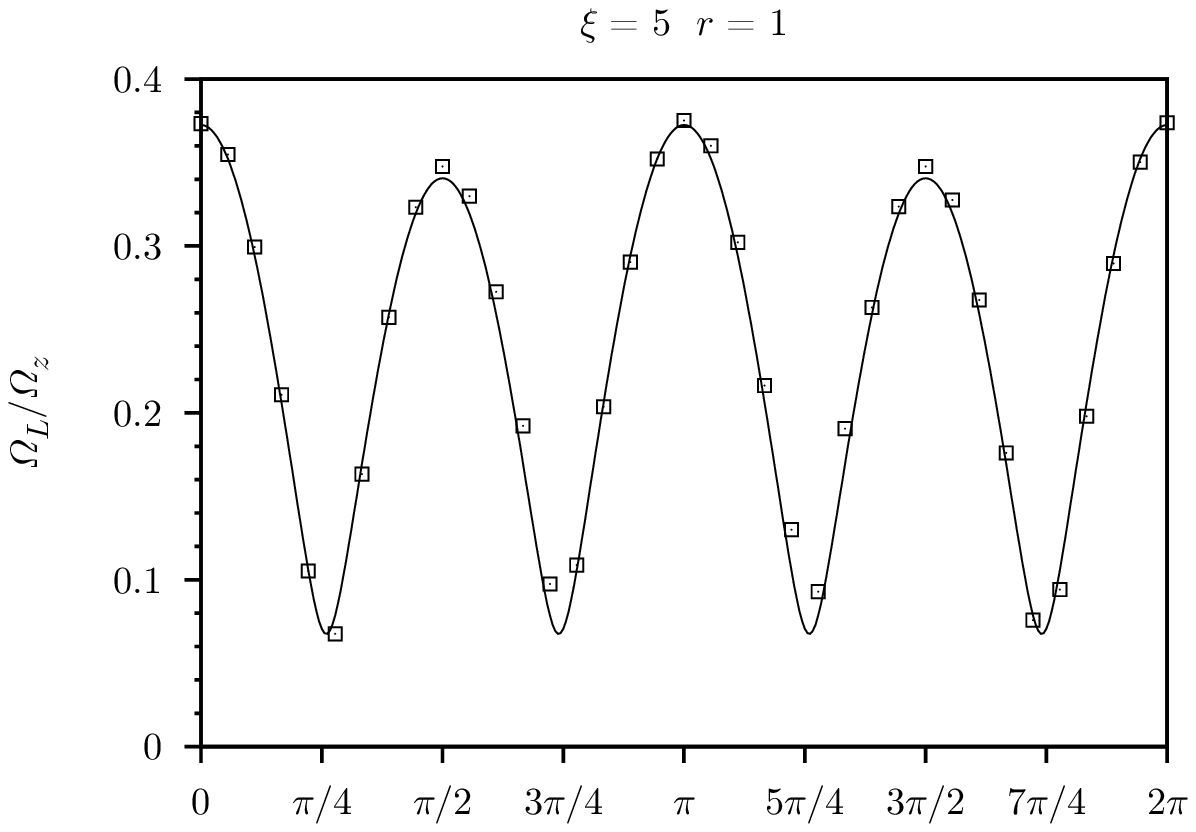}\\
  \includegraphics[width=8cm]{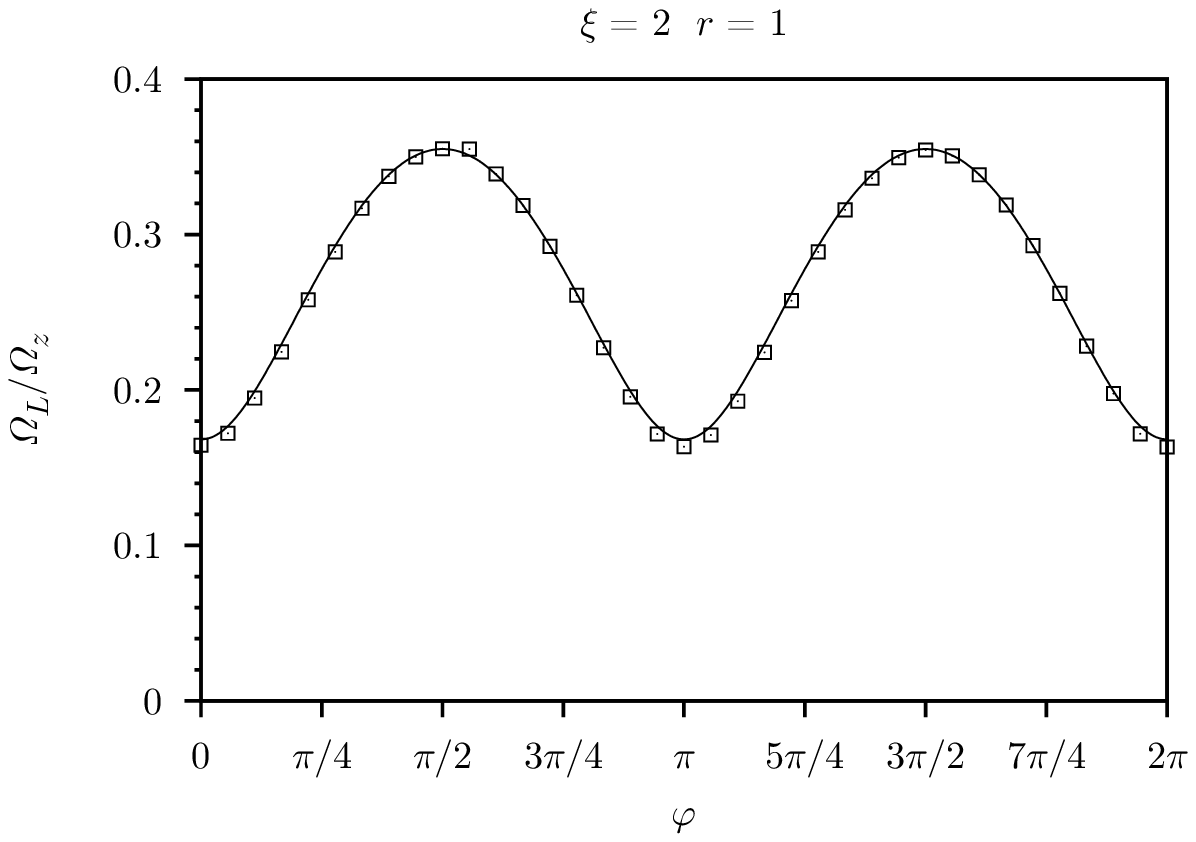}
  \caption{   Comparison  of   theoretical   predictions  (line)   and
    experimental points for $k=2$ and  different values of $\xi$ and $r$
    for the  ratio between the dressed  and unperturbed precession
    frequencies  as a  function of  the relative  phase between  the two
    Fourier components of the oscillating field (see eq.~\eqref{eq:anha:gen}).  }
  \label{fig:k2:x5x2_r1r1_comp}
\end{figure}

One fact which  stands out among the most  interesting features of the
system  considered  is that  the  non-linear  nature  of the  observed
phenomenon makes relevant the relative phase of the Fourier components
of    the   oscillating    field.    Figs.~\ref{fig:k3:x3x3_r1r2_comp}
and~\ref{fig:k2:x5x2_r1r1_comp}   are   devoted   to   comparing   the
dependency  of  $F$ on  one of those  parameters.   For a  large  enough
amplitude  of both  the  components, phase  values  exist causing  the
precession frequency to vanish.

Two comparisons for  $k=3$ are made in Fig.~\ref{fig:k3:x3x3_r1r2_comp},
where differently located  zeroes and local maxima of  $F$ appear. It is
worth noting that, as confirmed by  the experiment, in the case of $k=3$
the dependence of the relative  phase is symmetric with respect to the
$\varphi=\pi$ value.

Fig.~\ref{fig:k2:x5x2_r1r1_comp} reports  further comparisons made for
$k=2$.   In  these   cases  due  to  the  small   value  of  $r$,  the
interference-like pattern does not produce zeroes, while the number of
local maxima is  clearly dependent on the absolute  amplitude $\xi$. For
this value  of $k$, both  the model and  the experiment show  that the
dependence  of the  relative  phase has  an  additional symmetry  with
respect to the $\varphi=\pi/2$ value.

The  symmetries pointed  out in  the Figs.~\ref{fig:k3:x3x3_r1r2_comp}
and~\ref{fig:k2:x5x2_r1r1_comp},  can  be   interpreted  in  terms  of
properties of the Bessel function, according to the analysis presented
in  brief in  the  Appendix \ref{sec:app:prop:F}.   As  shown in  that
appendix, the symmetry around the $\pi$ value is a consequence of a $2
\pi$
periodicity occurring for both  odd and even values of $k$, while
the additional symmetry around $\pi/2$ occurs only for the even values
of $k$.

\begin{figure}[htbp]
  \centering
  \includegraphics[width=8cm]{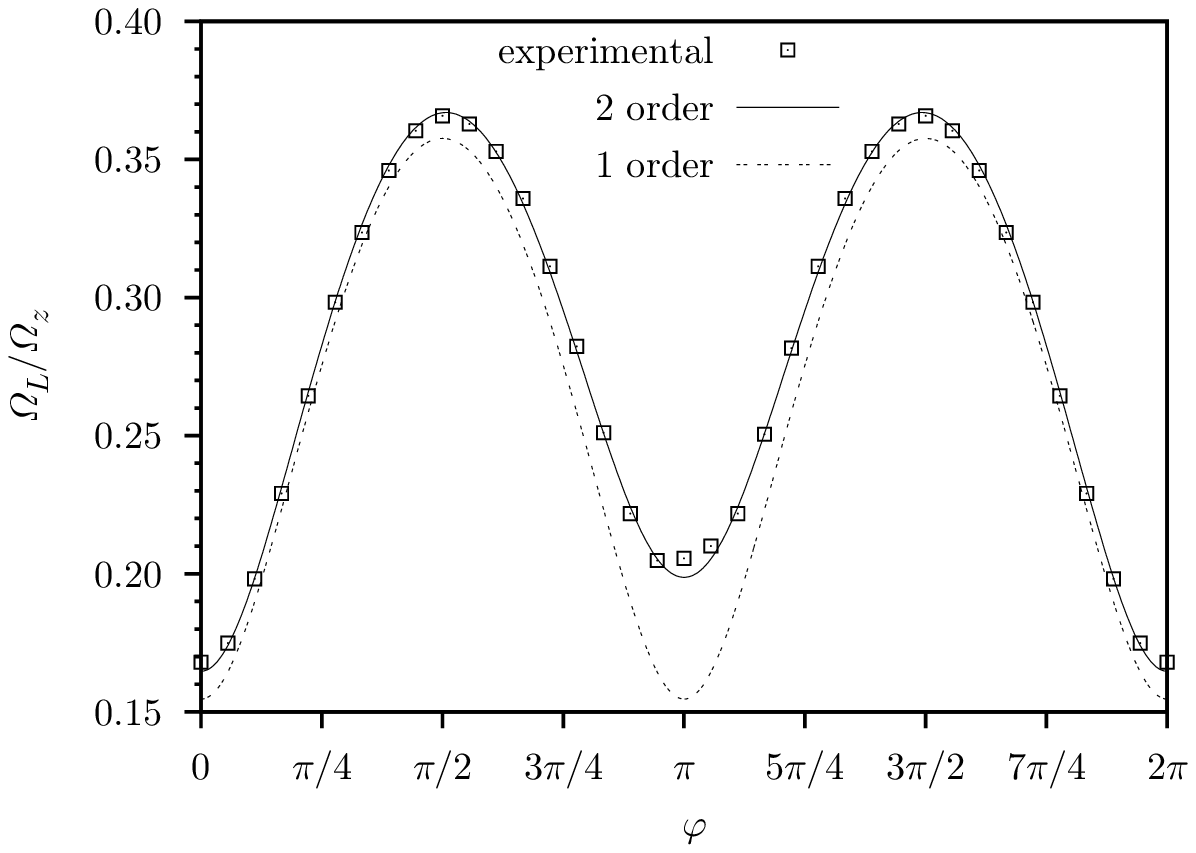}\\
  \includegraphics[width=8cm]{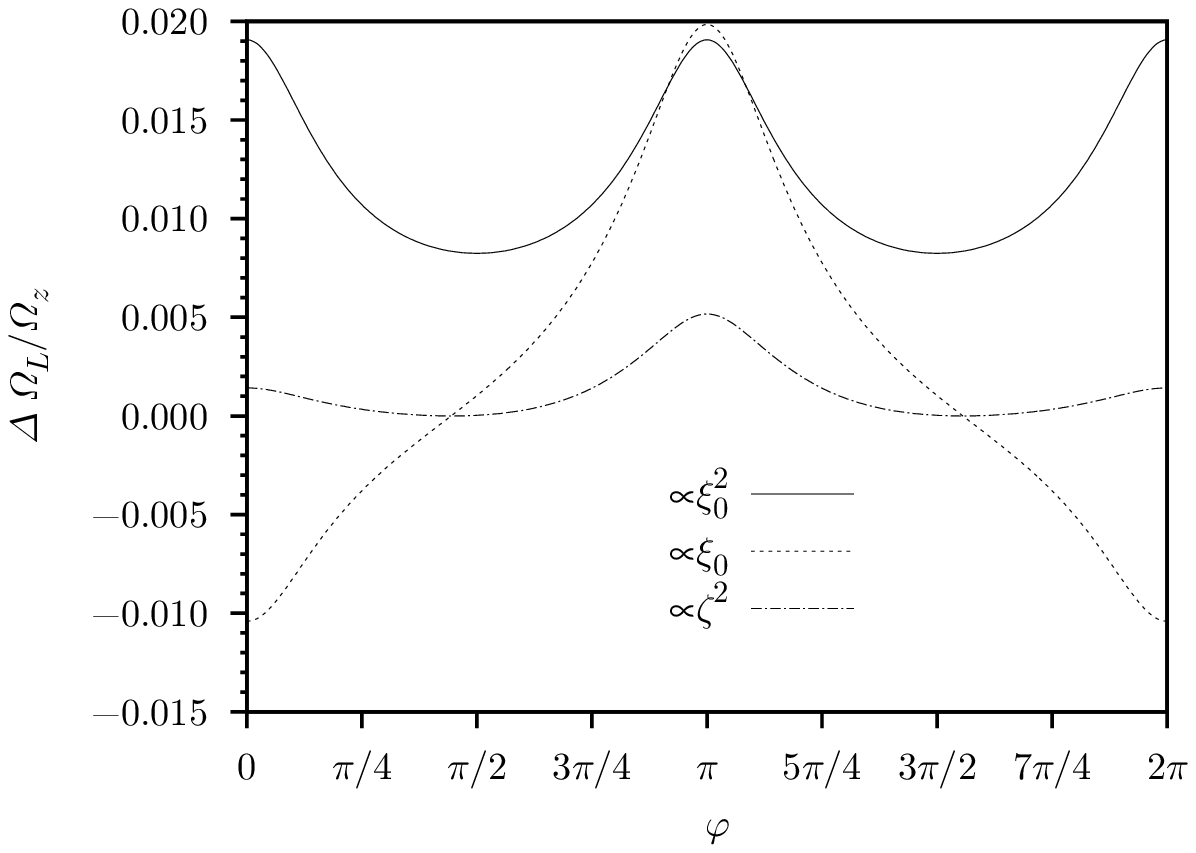}
  \caption{ (Upper)  Comparison of theoretical  predictions (line) and
    experimental points  in presence of a small  dc field superimposed
    to  the   ac  one.   The  parameters  are   $\xi_0  \equiv  \gamma
    B_x^{\mathrm{(dc)}}/\omega = 1.3 \cdot 10^{-2}$,  $\xi = 2$ , $r =
    1$, $k  = 2$ and $\zeta =  0.14$.  The solid line  is calculated using
    eq.~\eqref{eq:WL:1e2}     while    the     dashed     one    using
    eq.~\eqref{eq:dress:larmor}.  It is  clearly visible the effect of
    a  small  $\xi_0$.  (Lower)  The three  second  order  corrections
    according   to   eq.~\eqref{eq:WL:1e2}.   Notice  how   the   term
    proportional to  $\xi_0^2$ has the same value  for $\varphi=0$ and
    $\varphi=\pi$ and thus it is not responsible of the asymmetry.  }
\label{fig:x0}
\end{figure}

The even-$k$  symmetry can  be broken by  the presence of  a vectorial
perturbation along the oscillation  axis. In particular, a weak static
magnetic field  component added to  the alternating one results  in an
asymmetric curve
\begin{equation}
  \label{eq:simm:k:break}
  w_0\equiv \frac{\Omega_L}{\Omega_z}\bigg|_{\varphi=0} \neq
\frac{\Omega_L}{\Omega_z}\bigg|_{\varphi=\pi} \equiv w_\pi
\end{equation}
as  can be  seen  in Fig.~\ref{fig:x0},  where  it is  also shown  the
theoretical simulations with and  without the second order corrections
(see  eq.~\eqref{eq:WL:1e2}). A look  at the  right panel  reveals the
importance  of  the corrections  depending  of  the $G_n\,,\,n\neq  0$
coefficients through the  $c$ quantity (see eq.~\eqref{eq:chi:fin}) in
generating the asymmetry.

This asymmetry
has  potential applications  in designing  and setting  up apparatuses
like the magnetometer used in our experiment. The scalar nature of our
magnetometer  makes  it very  sensitive  to  variations  of the  field
component  parallel  to the  static  one  ($z$  component), while  the
transverse components  can be  detected and minimized  with procedures
responding  to the  second order  in $\xi_0$.   In  contrast, repeated
measurements  and  equalization of  $w_0$  and  $w_\pi$ constitute  an
effective  procedure  with a  first  order  response  in $\xi_0$  (see
eq.~\eqref{eq:WL:1e2}) for fine balancing the transverse dc component.
In this perspective, an  even-$k$ modulation technique may be proposed
to achieve a vector response of this type of scalar sensors.

It is  also worth nothing  that this asymmetry  cannot be seen  if the
modulation  is   purely  harmonic  (see   Appendix~\ref{sec:asym}  for
details),  at  least  in  the  perturbation order  considered  in  the
described model.

\section{Conclusions}
We  have studied  both  theoretically and  experimentally the  effects
produced  on the  Larmor precession  by a  transverse  magnetic field,
which oscillates at frequencies higher than the Larmor one, and has an
arbitrarily large  amplitude. The proposed  model is suitable  for the
interpretation  of  the   effect  of  a  non-harmonically  oscillating
transverse  field.  In  fact,  the nonlinear  nature  of the  system's
response makes  a harmonic  expansion unfeasible and  makes previously
proposed  analyses  (considering only  the  harmonic transverse  case)
unsuitable   for  describing   the  general   case  of   an  arbitrary
time-dependence of the transverse  magnetic field.  We have tested our
model experimentally,  studying in detail  the case of  a non-harmonic
transverse field constituted by  two Fourier components. The relevance
of the  relative phase of the  two components in  the determination of
the  observed  dressed Larmor  frequency  was  pointed  out.  We  also
demonstrated that  the non-harmonicity of the  alternating field gives
rise to a  first-order variation of the dressed  Larmor frequency with
respect to a small dc  field superimposed to the alternating one. This
observation  is of  interest  for its  potential  in designing  atomic
magnetometers with vectorial response.

\acknowledgments

The authors  thank Patricia H.  Robison for  reviewing the manuscript,
and Andrea Ruffini and Massimo  Catasta of GEM elettronica s.r.l., for
the valuable technical support.

\appendix

\section{Properties of $F$. }
\label{sec:app:prop:F}

In  this  appendix  we  list  some  properties of  the  quantity
\eqref{eq:larmor:ana:Omega}. An explicit expression can be obtained as
\begin{equation}
  \label{eq:app:F:expr:ini}
  F(\xi,\eta,k,\varphi) = \left[\rho_{k,0} + 2 \sum_{s \geq 1} \rho_{k,s} \, \cos(s \varphi )\right]^{1/2}
\end{equation}
where
\begin{equation}
  \label{eq:app:F:def:zks}
  \rho_{k,s} = \sum_{p=-\infty}^{\infty}
\bess{-k(p+s)}{\xi}
\bess{p+s}{\eta}
\bess{-kp}{\xi}
\bess{p}{\eta}.  
\end{equation}
Using the  property $J_{-m} =  (-1)^mJ_m$ of the Bessel  functions and
some straightforward manipulations one can see by inspection that
\begin{equation}
  \label{eq:app:F:zks:prop}
  \rho_{k,s} = (-1)^{(k+1)s}\rho_{k,s}
\end{equation}
which in the case  of odd $k$ reduces to a trivial  identity. For even $k$
it is  easily seen that  $\rho_{k,2n+1}\equiv 0 $ reducing  the $\varphi$
period in \eqref{eq:app:F:expr:ini} from $2\pi$ to $\pi$. This feature
is clearly visible in the Fig.~\ref{fig:k2:x5x2_r1r1_comp}.

\section{Asymmetry details}
\label{sec:asym}

Let's  consider the second  order corrections  to $\Omega_L$  which give
rise to the asymmetry. As can be seen from eq.~\eqref{eq:WL:1e2} one needs
to investigate the series \eqref{eq:alfa:beta}, which can be rewritten as
\begin{equation}
  \label{eq:a:b:app}
  a = \sum_{n=1}^{+\infty} \frac{G_0^*(G_n - G_{-n})}{n}
\qquad
b = \sum_{n=1}^{+\infty} \frac{|G_n|^2 - |G_{-n}|^2}{n} .
\end{equation}
Then let's evaluate these expressions for the case of a general harmonic
field as in eq.~\eqref{eq:harm:gen:int}.  Notice that in real experimental
situations  the  presence of  the  time  lag  $\tau_0$ is  unavoidable.  Thus we
assume that $\tau_0$ is a random parameter uniformly distributed
and  at  the  end of  the  calculations  all  the quantities  must  be
averaged.

 Using eq.~\eqref{eq:Gn:gen:harm} and the  property $J_{-n} = (-1)^n J_n$
of Bessel functions one arrives at
\begin{equation}
  \label{eq:a:b:harm}
  \mRe (a) = 2 \sum_{n=0}^{+\infty} \frac{J_0(\xi)J_{2n+1}(\xi)}{2n+1}\, \cos(2n+1)\tau_0
\qquad
b \equiv 0
\end{equation}
whose average
\begin{equation}
  \label{eq:media}
  \overline{\mRe (a)} \equiv \frac{1}{2\pi}\int_0^{2\pi} \mRe (a) \, \mathrm{d}\tau_0
\end{equation}
gives zero.  We can thus conclude that a purely harmonic signal cannot
give an asymmetric contribution.

The non-harmonic  case (see eq.~\eqref{eq:anha:gen}) on  the contrary shows
the asymmetry. In fact let the general non-harmonic signal be
\begin{equation}
  \label{eq:sign:anh:app}
  \phi(\tau)  = \cos(  \tau -  \tau_0) +  r \cos(  k(\tau -  \tau_0) +
  \varphi ),
\end{equation}
using the same device of eq.~\eqref{eq:Gn:anha:gen} we can write
\begin{equation}
  \label{eq:Gn:app}
  G_n(\tau_0)   =   \mexp{i\,\xi\,(\sin\tau_0   +  (r/k)\sin(k\tau_0   -
    \varphi) +(r/k)\sin\varphi )} \, \mexp{-i\,n\,\tau_0}\, G_n
\end{equation}
where the $G_n$ on the right-hand side can be 
read  from eq.~\eqref{eq:Gn:anha:gen}.   
From  this   result   easily  follows   that
\begin{equation}
  \label{eq:Gn:sq:app}
  |G_n(\tau_0)|^2 = |G_n|^2  
\end{equation}
showing  that the  leading  contribution to  the  Larmor frequency  is
unaffected  by  an  overall  phase. Moreover  $G_{-n}$  cannot  be simply
related to $G_n$ (as in  Bessel functions case), thus
the $b$ parameter in \eqref{eq:a:b:app} generally is non-zero.
For  the other
series after some algebra one finds
\begin{equation}
  \label{eq:a:anh:app}
  \mRe (a)  = \sum_{n=1}^{+\infty}\left\{  \frac{\mRe \left[ G_0^*(G_n  - G_{-n})
    \right] }{n} \cos n\tau_0
  - \frac{\mIm \left[ G_0^*(G_n - G_{-n})
    \right] }{n} \sin n\tau_0 \right\}
\end{equation}
whose average is zero.

\bibliographystyle{apsrev}
\bibliography{ref}

\begin{thebibliography}{11}
\expandafter\ifx\csname natexlab\endcsname\relax\def\natexlab#1{#1}\fi
\expandafter\ifx\csname bibnamefont\endcsname\relax
  \def\bibnamefont#1{#1}\fi
\expandafter\ifx\csname bibfnamefont\endcsname\relax
  \def\bibfnamefont#1{#1}\fi
\expandafter\ifx\csname citenamefont\endcsname\relax
  \def\citenamefont#1{#1}\fi
\expandafter\ifx\csname url\endcsname\relax
  \def\url#1{\texttt{#1}}\fi
\expandafter\ifx\csname urlprefix\endcsname\relax\def\urlprefix{URL }\fi
\providecommand{\bibinfo}[2]{#2}
\providecommand{\eprint}[2][]{\url{#2}}

\bibitem[{\citenamefont{Haroche et~al.}(1970)\citenamefont{Haroche,
  Cohen-Tannoudji, Audoin, and Schermann}}]{CCT_1970}
\bibinfo{author}{\bibfnamefont{S.}~\bibnamefont{Haroche}},
  \bibinfo{author}{\bibfnamefont{C.}~\bibnamefont{Cohen-Tannoudji}},
  \bibinfo{author}{\bibfnamefont{C.}~\bibnamefont{Audoin}}, \bibnamefont{and}
  \bibinfo{author}{\bibfnamefont{J.~P.} \bibnamefont{Schermann}},
  \bibinfo{journal}{Phys. Rev. Lett.} \textbf{\bibinfo{volume}{24}},
  \bibinfo{pages}{861} (\bibinfo{year}{1970}).

\bibitem[{\citenamefont{Kenkre and Raghavan}(2000)}]{Kenkre_art186}
\bibinfo{author}{\bibfnamefont{V.~M.} \bibnamefont{Kenkre}} \bibnamefont{and}
  \bibinfo{author}{\bibfnamefont{S.}~\bibnamefont{Raghavan}},
  \bibinfo{journal}{J. Opt. B: Quantum Semiclass. Opt.}
  \textbf{\bibinfo{volume}{2}}, \bibinfo{pages}{686} (\bibinfo{year}{2000}).

\bibitem[{\citenamefont{Holthaus and Hone}(1996)}]{Holtaus_1996}
\bibinfo{author}{\bibfnamefont{M.}~\bibnamefont{Holthaus}} \bibnamefont{and}
  \bibinfo{author}{\bibfnamefont{D.~W.} \bibnamefont{Hone}},
  \bibinfo{journal}{Phil. Mag. B} \textbf{\bibinfo{volume}{74}},
  \bibinfo{pages}{105} (\bibinfo{year}{1996}).

\bibitem[{\citenamefont{Dunlap and Kenkre}(1986)}]{Kenkre_art81}
\bibinfo{author}{\bibfnamefont{D.~H.} \bibnamefont{Dunlap}} \bibnamefont{and}
  \bibinfo{author}{\bibfnamefont{V.~M.} \bibnamefont{Kenkre}},
  \bibinfo{journal}{Phys. Rev. B} \textbf{\bibinfo{volume}{34}},
  \bibinfo{pages}{3625} (\bibinfo{year}{1986}).

\bibitem[{\citenamefont{Eckardt et~al.}(2009)\citenamefont{Eckardt, Holthaus,
  Lignier, Zenesini, Ciampini, Morsch, and Arimondo}}]{PhysRevA.79.013611}
\bibinfo{author}{\bibfnamefont{A.}~\bibnamefont{Eckardt}},
  \bibinfo{author}{\bibfnamefont{M.}~\bibnamefont{Holthaus}},
  \bibinfo{author}{\bibfnamefont{H.}~\bibnamefont{Lignier}},
  \bibinfo{author}{\bibfnamefont{A.}~\bibnamefont{Zenesini}},
  \bibinfo{author}{\bibfnamefont{D.}~\bibnamefont{Ciampini}},
  \bibinfo{author}{\bibfnamefont{O.}~\bibnamefont{Morsch}}, \bibnamefont{and}
  \bibinfo{author}{\bibfnamefont{E.}~\bibnamefont{Arimondo}},
  \bibinfo{journal}{Phys. Rev. A} \textbf{\bibinfo{volume}{79}},
  \bibinfo{pages}{013611} (\bibinfo{year}{2009}),
  \urlprefix\url{http://link.aps.org/doi/10.1103/PhysRevA.79.013611}.

\bibitem[{\citenamefont{Ghosh}(1999)}]{1999_ghosh}
\bibinfo{author}{\bibfnamefont{G.}~\bibnamefont{Ghosh}}, \bibinfo{journal}{Opt.
  Comm.} \textbf{\bibinfo{volume}{163}}, \bibinfo{pages}{95}
  (\bibinfo{year}{1999}).

\bibitem[{\citenamefont{{Johnson} et~al.}(2010)\citenamefont{{Johnson},
  {Schwindt}, and {Weisend}}}]{johnson2010}
\bibinfo{author}{\bibfnamefont{C.}~\bibnamefont{{Johnson}}},
  \bibinfo{author}{\bibfnamefont{P.~D.~D.} \bibnamefont{{Schwindt}}},
  \bibnamefont{and}
  \bibinfo{author}{\bibfnamefont{M.}~\bibnamefont{{Weisend}}},
  \bibinfo{journal}{Applied Physics Letters} \textbf{\bibinfo{volume}{97}},
  \bibinfo{pages}{243703} (\bibinfo{year}{2010}).

\bibitem[{\citenamefont{Blanes et~al.}(2009)\citenamefont{Blanes, Casas, Oteo,
  and Ros}}]{2009_Blanes}
\bibinfo{author}{\bibfnamefont{S.}~\bibnamefont{Blanes}},
  \bibinfo{author}{\bibfnamefont{F.}~\bibnamefont{Casas}},
  \bibinfo{author}{\bibfnamefont{J.}~\bibnamefont{Oteo}}, \bibnamefont{and}
  \bibinfo{author}{\bibfnamefont{J.}~\bibnamefont{Ros}},
  \bibinfo{journal}{Physics Reports} \textbf{\bibinfo{volume}{470}},
  \bibinfo{pages}{151 } (\bibinfo{year}{2009}), ISSN \bibinfo{issn}{0370-1573},
  \urlprefix\url{http://www.sciencedirect.com/science/article/pii/S03701573080%
04092}.

\bibitem[{\citenamefont{Abramowitz and Stegun}(1964)}]{abramowitz:stegun}
\bibinfo{author}{\bibfnamefont{M.}~\bibnamefont{Abramowitz}} \bibnamefont{and}
  \bibinfo{author}{\bibfnamefont{I.~A.} \bibnamefont{Stegun}},
  \emph{\bibinfo{title}{Handbook of Mathematical Functions with Formulas,
  Graphs, and Mathematical Tables}} (\bibinfo{publisher}{Dover},
  \bibinfo{address}{New York}, \bibinfo{year}{1964}), \bibinfo{edition}{ninth
  dover printing, tenth gpo printing} ed., ISBN \bibinfo{isbn}{0-486-61272-4}.

\bibitem[{\citenamefont{Dattoli and Torre}(1996)}]{Dattoli1996}
\bibinfo{author}{\bibfnamefont{G.}~\bibnamefont{Dattoli}} \bibnamefont{and}
  \bibinfo{author}{\bibfnamefont{A.}~\bibnamefont{Torre}},
  \emph{\bibinfo{title}{Theory and Applications of Generalized Bessel
  Functions}} (\bibinfo{publisher}{Aracne Editrice, Rome},
  \bibinfo{year}{1996}).

\bibitem[{\citenamefont{{Ito} et~al.}(2003)\citenamefont{{Ito}, {Shimomura},
  and {Yabuzaki}}}]{ito03}
\bibinfo{author}{\bibfnamefont{T.}~\bibnamefont{{Ito}}},
  \bibinfo{author}{\bibfnamefont{N.}~\bibnamefont{{Shimomura}}},
  \bibnamefont{and}
  \bibinfo{author}{\bibfnamefont{T.}~\bibnamefont{{Yabuzaki}}},
  \bibinfo{journal}{Journal of the Physical Society of Japan}
  \textbf{\bibinfo{volume}{72}}, \bibinfo{pages}{1302} (\bibinfo{year}{2003}).

\end{thebibliography}
\end{document}